# Spin Coated Plasmonic Nanoparticle Interfaces for Photocurrent Enhancement in Thin Film Si Solar Cells

Miriam Israelowitz,[†] Jennifer Amey,[*] Tao Cong,[*] and Radhakrishna Sureshkumar[*‡]

[†]*Department of Electrical Engineering, Syracuse University, Syracuse, NY 13244*

[*]*Department of Biomedical and Chemical Engineering, Syracuse, NY 13244*

[‡]*Email: rsuresk@syr.edu*

September 5, 2013

**Abstract**

Nanoparticle (NP) arrays of noble metals strongly absorb light in the visible to infrared wavelengths through resonant interactions between the incident electromagnetic field and the metal's free electron plasma. Such plasmonic interfaces enhance light absorption and photocurrent in solar cells. We report a cost effective and scalable room temperature/pressure spin-coating route to fabricate broadband plasmonic interfaces consisting of silver NPs. The NP interface yields photocurrent enhancement (PE) in thin film silicon devices by up to 200% which is significantly greater than previously reported values. For coatings produced from Ag nanoink containing particles with average diameter of 40 nm, an optimal NP surface coverage $\phi$ of 7% was observed. Scanning electron microscopy of interface morphologies revealed that for low $\phi$, particles are well-separated, resulting in broadband PE. At higher $\phi$, formation of particle strings and clusters caused red-shifting of the PE peak and a narrower spectral response.

# 1 Introduction

Thin film silicon (Si) solar cells reduce material cost of photovoltaic (PV) systems and offer a means to more affordable renewable energy production. Thin film silicon PVs have an average cell thickness of 300-500 nm as compared to 200-500 μm for bulk crystalline Silicon (c-Si) ones [1]. However, the major disadvantage of thin film Si PVs is the relatively low efficiency of light to power conversion due to the low light absorption rate of Si and the reduced optical path length of the thin film. Currently, the best research cell efficiencies recorded for thin-film single junction amorphous Si is 13.4% compared with bulk c-Si of 27.6% [2]. Therefore, in order to realize the potential of thin film Si PVs, efficient broadband light trapping technologies need to be integrated into the device design.

The use of plasmonic interfaces offer much promise for increasing the light trapping efficiency of thin film PVs. Plasmonics play a large role in determining the optical properties of metal nanoparticles (NPs). Specifically, the phenomenon of localized surface plasmon resonance (LSPR) which refers to the strong coupling of the incident electromagnetic field and the free electron plasma of the metal determines the frequencies at which metallic NPs strongly scatter light. Hence, metallic NP coatings can be tailored to create plasmonic interfaces on Si thin film PVs to efficiently scatter light at large angles into the underlying semiconducting layer and increase the optical path length. There are two fundamental mechanisms proposed to explain how the NPs lengthen the optical path in thin film PVs. First, when the NP diameter is significantly smaller than the wavelength of the incident light, the optical field characteristics of the particle can be approximated as a point dipole, which can re-radiate the light acting as a powerful scattering element [3], [4]. For a metal NP with complex permittivity $\varepsilon_p$ embedded in a homogenous medium with permittivity $\varepsilon_m$, depending on the particle volume V and the incident wavelength λ the effective scattering cross section can be much larger than the physical cross section of the particle. The scattering and absorption cross sections, $C_{scat}$ and $C_{abs}$ respectively, are given by [3], [4]:

$$C_{scat} = \frac{1}{6\pi}\left(\frac{2\pi}{\lambda}\right)^4 |\alpha|^2, \quad C_{abs} = \frac{2\pi}{\lambda}\text{Im}[\alpha] \qquad (1)$$

where

$$\alpha = 3V\left[\frac{\varepsilon_p/\varepsilon_m - 1}{\varepsilon_p/\varepsilon_m + 2}\right] \qquad (2)$$

is the polarizability of the particle. The bulk plasmon resonance occurs when $\varepsilon_p = -2\varepsilon_m$ for which the scattering cross sections increase in size dramatically [3]. Second, a dipole nearby a planar interface of higher optical density can strongly couple evanescent waves otherwise lost, resulting in near field light concentration [5]. A dipole can be modeled as the superposition of its propagating and evanescent waves [5], [6]. Propagating waves transmitted into the dielectric will do so at angles less than the critical angle of the dielectric boundary. Conversely, evanescent waves transmitted into a dielectric will propagate at angles greater than the critical angle [7]. The power of the radiation from evanescent sources decreases exponentially depending on the distance of the dipole from the surface. The scattering of light by a NP shows a symmetric radiation pattern when embedded in a homogenous material. However, this pattern changes when brought into the vicinity of a dielectric surface, in which the light will scatter primarily into the dielectric of larger permittivity [3]. The fraction of incident light scattered into the substrate using a dipole near a dielectric interface could be as large as 96% [3].

Light trapping and photocurrent enhancement (PE) by plasmonic interfaces have been well documented in literature. There have been several methods reported in the literature to create plasmonic interfaces, such as electron beam lithography [8], thermal evaporation [9], nanoimprinting lithography [10], and ns and fs pulsed laser patterning of ultra-thin films [11-13]. The most common plasmonic interfaces used for light trapping consists of NP islands formed through thermal evaporation of a metal film followed by annealing [8], [14], [15]. However, this method creates typically a limited range of particle sizes and shapes. Conversely, synthesis of plasmonic NPs using colloid chemistry techniques offers much greater control over particle size, shape, mono-dispersity, and passivating shell layers [16], [17]. Subsequently, manufacturing of plasmonic interfaces from NP suspensions using wet chemistry methods could offer robust and controllable means for tailoring plasmonic interfaces with desired optical properties [18], [19]. Herein, we report a cost-effective and scalable room temperature/pressure nanomanufacturing process based on a spin-coating technique to create broadband light trapping plasmonic interfaces on silicon-on-insulator (SOI) devices. Such interfaces are shown to facilitate strong PE of the SOI device up to 200%.

## 2  Methods

**General**

All chemicals were bought from Aldrich and used as received unless otherwise stated. A tabletop centrifuge (Eppendorf 5418) was used for NP purification and isolation. Ocean Optics UV/Vis spectrometer (USB4000- UV-VIS) was used to characterize NP absorption spectra. Particle size was measured using dynamic light scattering (DLS) on a Malvern Zetasizer Nano Series instrument utilizing 173° backscatter. SOI wafers were purchased from University Wafer. The device contacts were fabricated using a thermal evaporator (CVC SC4500) and SEM images (Zeiss Ultra 55) were taken at the Cornell NanoScale Science & Technology Facility (CNF). NPs were applied to SOI devices using a Laurell Technologies Corporation Spin Coater (WS-650Mz-23NPP).

**Preparation of SOI devices**

SOI cells were used to measure the NP effects on photocurrent. SOI cells are simple devices that have an accessible open surface that enabled easy monolayer coverage of NPs close to the c-Si active layer within a few nanometers. The wafer was cleaned using the RCA process and metal finger contacts were deposited by thermal evaporation. A mask was used to add 1mm x 10mm aluminum contact pairs with a finger distance of 1 mm apart using thermal vapor deposition and the contacts were deposited to a thickness of 35nm. The wafer was allowed to rest for a minimum of four hours in standard atmospheric conditions to let the native $SiO_2$ regrow on the top active layer. The wafer was a bonded c-Si n-type wafer, doped with Phosphorous and a resist of 1-10 $\Omega$-cm. In true thin film a-Si solar cells the a-Si active layer is thinner than many commercially available SOI wafers, however SOI wafers are thin enough to approximate the actual active device thickness. In our case the c-Si SOI active layer was ~3-4x thicker than the active layer of standard thin film solar cells.

**Preparation of silver ink solutions**

A series of ethanol solutions containing ethylene glycol capped Ag NPs (10, 5, 1, 0.5, 0.2, 0.1, 0.05, 0.01% w/v) were prepared from a stock solution of the exact same materials (20% w/v). DLS measurements showed an average particle size, $d=40 \pm 1$ nm. Ethanol was chosen to dilute the Ag NPs because of the relatively high volatility of the liquid, and would dry off of a surface quickly after being deposited without leaving residues that may block light and suppress current.

**Synthesis of glucose capped silver NPs**

Size controlled NPs may be produced through chemical synthesis by varying the pH of solution during production [20]. An aqueous solution of colloidal Ag particles stabilized by glucose was prepared by adding ammonia (0.02 mol L$^{-1}$) to Silver nitrate solution (10$^{-3}$ mol L$^{-1}$). Thereafter, a solution of glucose (0.01 mol L$^{-1}$) was added and the reaction was stirred vigorously for 2 minutes. DLS measurements showed an average NP size, $d = 31 \pm 1$ nm. In a separate experiment to make larger particles, the procedure was repeated as stated above, but the ammonia concentration was increased to (0.1 mol L$^{-1}$) and the reaction incubation time was increased to 12 minutes. DLS measurements showed an average NP size, $d = 69 \pm 1$ nm. The particles were cleaned by centrifuging the solution for ten minutes to excess glucose.

**Preparation of plasmonic interfaces**

In all experiments, the bare SOI device was placed into a spin coater and vacuum applied. The speed of the spin coater was ramped from 0 rpm to 8000 rpm over a time period of 0.5 minutes. During the spinning speeds of 2000-8000 rpm, a solution of colloidal particles (300 μl) was deposited onto the bare SOI device using an eppendorf micropipette, one drop at a time in a continual fashion. After the total volume of solution was deposited, the coated SOI device was left to spin at 8000 rpm and dried under vacuum for an additional 29.5 minutes.

**Photocurrent measurements**

The SOI device was a metal-semiconductor-metal (MSM) photodetector that has a single doped n-type c-Si layer with lateral Schottky barrier contacts as shown in Figure 1 (a), similar to experimental set-up reported by Pillai *et al*. [14]. An electrical gradient was applied to the contacts to induce the "on" state of the diode. The MSM photodetector is favorable for photocurrent measurement experiments because the photocurrent generated is linearly proportional to the optical power of the incident light [21-23]. Photocurrent is a key variable in determining both the open current voltage and the short circuit current of a solar cell. Improving the photocurrent generated would translate into greater solar cell efficiency. A tunable step-motor monochromator irradiated the light at discrete wavelengths from 400 to 1000 nm as seen in Figure 1 (b). The light source was a halogen lamp under non-AM 1.5 standard conditions. The irradiance larger than 1000 nm had very low power, and thus was difficult to discern from noise. The beam was directed to an optical microscope and focused on the sample and reference using an internal beam splitter. The induced photocurrent was measured via probes using a pre-amplifier across the contacts. The signal was extracted using a lock in amplifier and sent to the control computer. The short circuit current was recorded as a function of wavelength over the contact area using the analysis computer. The photocurrent response for the SOI device was recorded several times on each sample at different points on the p-n metal contacts and averaged to get the photocurrent response. The PE is defined as the ratio of the difference between the photocurrent generated after and before NP deposition to that before deposition expressed in terms of a percentage [14-15]. Therefore, the uncoated SOI device was first measured to get the relative photocurrent. Immediately after the bare surface measurement, the Ag nano ink or synthesized particle solutions were deposited via spin-coating and dried at ambient conditions.

As little time as possible was left between the measurements to be sure the collector time variation was small.

## 3  Results

**Silver nano ink**

The samples used to measure the PE caused by the silver (Ag) nano ink coating were prepared on SOI wafers with a 2 μm n-type crystalline silicon (c-Si) active layer by spin coating. A schematic depicting the structure of the SOI device is shown in Figure 1 (a). The structure of the wafer consisted of a neutral Si substrate of 625 μm thickness, a buried oxide layer of 1μm thickness, and a top c-Si n-type layer of 2 μm thickness. The samples were irradiated at discrete wavelengths and the resulting photocurrent was measured and recorded as shown in Figure 1(b). The absorption range for the 2 μm thick SOI device was 400-1000 nm. All samples showed the same shifted interference fringe pattern at wavelengths larger than 700 nm superimposed on any bulk plasmon resonance peaks. Interference patterns appear in the raw data because of transmission and reflection interference within the thin c-Si active layer. These fringes are phase shifted by the silver particles so that the photocurrent responses show the constructive or destructive PE patterns [4], [24].

Figure 2 shows the relative photocurrent enhancement responses for eight different Ag NP concentrations. For Ag NP ink concentrations in the range of 0.01-0.1% w/v, the resulting PE curves show a significant broadband response. According to Equations (1) and (2), the bulk plasmon resonance peak of silver in air occurs at $\lambda$ = 375 nm which is outside the range of operation of our device. Two small peaks of increasing strength with increasing concentration appear at $\lambda \approx$ 475 and 760 nm because of the long range dipole-dipole interactions near a dielectric surface [25-26]. As the particle coverage becomes denser, the distance between particles decreases and long range interactions increase. For Ag NP ink concentrations in the range of 1-10% w/v, the recorded PE curves show substantial red-shifting. For λ between 400 to 525 nm, PE decreases, but it quickly increases for λ > 525 nm. Because the curve cannot be shifted past 1100 nm which corresponds to the bandgap of c-Si, red-shifting of the bulk plasmon resonance results in the observed narrowing of the spectrum. The greatest overall PE is observed for coatings fabricated from a nano-ink solution with a concentration ρ = 0.1% w/v of Ag, for which the spectrum exhibits the greatest overall broadband response. The *overall PE* is defined as the ratio of the difference between the integrals of the photocurrent response curve after and before NP deposition to that before NP deposition in terms of a percentage. PE increases with increasing ρ until up to ρ = 0.1% w/v. In this concentration regime, the coating consists of well-separated particles: see Figure 5 and discussion below. Therefore, the total scattered flux, and hence the PE, can be expected to increase linearly with surface coverage. This is corroborated by the observed approximately linear increase in PE with increasing ρ for ρ ≤ 0.1% w/v, as shown in Figure 3. Moreover, for ρ ranging between 0.1 and 0.5% w/v, the PE is relatively insensitive to ρ and a near-plateau region can be seen. Further increase in ρ results in a substantial reduction in PE. This observed behavior in PE can be understood based on the changes in the coating morphology as a function of ρ: see the "Discussion" section below.

**Effect of Particle Size: Glucose capped silver NPs**

It is difficult to obtain exquisite control of particle size in commercially available Ag nano-ink solutions. Hence, in order to decipher the role of particle size on PE, we synthesized spherical Ag NPs encapsulated within a glucose shell following the procedure outlined in the "Methods" section. Two separate batches of Ag NPs were synthesised with average diameters *d* of 31 ± 1 nm and 69 ± 1 nm, as determined from dynamic light scattering (DLS) measurements. The smaller particle batch had a glucose shell of ~1 nm whereas the larger particle batch had an ~20 nm glucose shell. The particle concentration was 0.002% w/v. The SOI device used for photocurrent measurements had a 1.5 μm thick c-Si active layer. The particles were deposited via spin-coating and the photocurrent response measured in the exact same manner as for the Ag nano ink coatings. Figure 4 shows the photocurrent response for (a) 31 nm and (b) 69 nm particles. The overall PE for the 31 nm particles was 49 ± 1 %. The PE obtained for $d$ = 69 nm was 199 ± 3 %. This large PE can be explained based on the quadratic increase in the scattering cross section with increasing particle volume. The albedo of particles at 500 nm wavelength is shown in Figure 4(c) (blue line). The albedo $A$ is defined as $C_{scat}/(C_{scat} + C_{abs})$. The normalized PE responses of the two particle sizes tested are shown in red circles in Figure 4(c) follow the same trend as $A$, suggesting that scattering from the dipoles (particles) is the key mechanism of PE in dilute systems.

## 4 Discussion

In this section, we correlate the percentage surface coverage ϕ and surface morphology of the plasmonic interfaces fabricated by spin-coating to the particle concentration ρ in the feed solution. Further, we utilize theoretical models for the optical response of particle strings and clusters to understand the observed PE-morphology relationship [27], [28]. The effectiveness of NP-mediated scattering mechanisms on PE depends on a number of factors such as the NP size, shape, material, particle-substrate distance, and surface morphology/coverage.

The morphological features and ϕ of the coatings were determined by analyzing the scanning electron microscope (SEM) images of the interface. Figure 5 (a) shows the SEM images for coatings created from Ag nano-ink solutions with $\rho$ = 0.1, 1, 5, and 10% w/v. As seen in Figure 2, the greatest overall PE occurs for $\rho$ = 0.1% w/v ($\phi$ = 7%). For low $\phi$, as shown in Figure 5 (a *i*), the particles are well-separated and the scattering cross section can be approximated by Eqs. (1) and (2). Further, since there is limited interaction among the particles, the total scattered flux will increase linearly with increasing $\phi$. However, at higher particle concentrations, particles tend to aggregate. This aggregation can be understood based on the dynamics of the spin-coating process. Spin-coating results in the formation of a thin layer of fluid in which particles are embedded. As the fluid film evaporates, attractive capillary forces can pull the particles closer to one another [29]. As the particle concentration increases, the initial inter-particle separation in the fluid film decreases and within the time scale of drying of the fluid film, particles can attain maximal proximity, i.e., the inter-particle distance becomes close to particle diameter.

The optical properties of clusters and strings can be understood by examining the inter-particle distances. When two metallic particles are within close proximity of each other, each individual particle unit forms a charge distribution of end-to-end dipoles [27], [28]. Linear chains of NPs act as a single dipole and the resultant effect is a red-shifting of the bulk plasmon resonance, which can be explained by the hybridization theory [27], [28]. Specifically, the conducting electrons in the metal NPs may be modeled as a hybrid fluid form of electron density with a unique dielectric constant. As the inter-particle distance decreases, the hybridization effect and the red-shifting of the bulk resonance plasmon mode become more pronounced.

When particles are not in a linear formation, the dipoles no longer align end-to-end. Instead, the inter-particle interactions arise from multiple spatial directions and multi-poles are induced within the individual particles. Multi-poles are inefficient radiators, and have much lower absorption and scattering strengths compared to single particles [27], [30]. Multi-poles also induce a separate bulk plasmon resonance peak that is different from the dipole peak, which becomes more distinct and disperse as more particles are added to the cluster. Clusters are extremely sensitive to light polarization and geometric arrangements. Super-radiant (dipole) and sub-radiant (multi-pole) modes form separate resonance peaks that quickly separate and diminish in intensity as the cluster gets larger because of destructive interference. The super-radiant plasmon mode always red-shifts, while the sub-radiant modes either show a slight-blue shift or no shift at all [27], [28]. This effect is seen in both aggregates of NPs in solution and also in large, extended close packed disordered and ordered NP interfaces [27].

Particle aggregates on the samples can be seen in Figure 5 (a *ii-iv*) for $\rho \geq 1\%$. They are classified into strings (red) and clusters (yellow) and highlighted in Figure 5 (b). A string is defined to be an approximately linear aggregate of three or more particles. A cluster is defined to be a non-linear aggregate of three or more particles where each particle has two or more neighbors within a single particle diameter. In Figure 2, the PE for $\rho = 0.1\%$ w/v shows a uniform broadband response. The only noticeable peaks are observed in the PE spectrum are resonances from Ag in $SiO_2$ and Si. As the coatings become denser and particle aggregation occurs, the PE spectrum exhibits shifting of responses into the red band. The red-shifting narrows the overall bandwidth of PE since there is no absorbance above the bandgap wavelength of c-Si. The net result of the narrowing of the bandwidth is a reduction in PE. Irrespective of the morphology of the aggregates, i.e., whether they are strings or clusters, they exhibit strong multi-pole characteristics, thus reducing the overall enhancement. As strings get longer and clusters get larger, the red shifting becomes more pronounced as the bulk plasmon resonance peak bandwidths narrow, with the 10% w/v solution producing the largest red shift. Figure 6 shows that ϕ as well as the number of strings and clusters as a function of particle ρ in the feed solution. The surface coverage first increases linearly with increasing ρ and appears to approach a plateau for higher ρ possibly due to the formation of multilayered aggregates. The number of string and clusters grows linearly with increasing ρ for ρ ≤ 1% w/v. As ρ is further increased, the rate of formation of aggregates decreases at the expense of having larger clusters and strings. This results in the saturation in the number of strings and clusters for ρ between 5% and 10% solution w/v. The larger aggregates are detrimental to PE as seen in Figure 3.

The 31 nm glucose-capped particles showed a broadband enhancement with modest enhancement gains. The maximum enhancement gain using thermal evaporation to create particle sizes of 16 nm is 33%, where the particles are closely packed Ag disks [14]. This shows that using chemical fabrication and the spin-coating technique produces superior PE results compared to the thermal evaporation method. Chemically synthesizing the NP to more desirable shapes such as a cylinder may produce an even greater PE response [8]. The 69 nm particles produced huge PE gains corresponding to the greater dipole strength. The increasing particle volume also causes red-shifts in the bulk plasmon resonances. As the particle size increases, multi-pole modes contribute to the absorption and scattering cross section[31]. In Figure 4 (b), four peak enhancements, namely at ≈490, 575, 710 and 830 nm range can be seen. The peaks at 490 and 830 nm can be attributed to the long range dipole-dipole interactions. Due to particle size, the long range interaction resonance peaks become slightly red shifted as compared to their smaller diameter counterparts [26]. Additionally, the peaks occurring at ≈575 and 710 nm is the signature of the Ag-glucose core-shell architecture. Because of the relatively larger particle size compared to the visible wavelength range, Equations (1) and (2) do not accurately capture the scattering and absorption cross sections. Hence, Mie theory was used to calculate the scattering and absorption cross sections of a 69 nm Ag nanoparticle embedded in glucose [17, 32]. As shown in Figure 7, two bulk plasmon resonance peaks are predicted at 560 and 730 nm, corresponding to the multi-pole and dipole resonances respectively, consistent with the peaks in the measured photocurrent response.

Chemically synthesized silver nanoparticles have been spin-coated onto a SOI device to fabricate light harvesting plasmonic interfaces, resulting in PE of up to 200%. The correlation between PE and interface morphology has been identified. Specifically, we have shown the existence of an optimal window in the surface coverage of the particles (~7%) for which PE is maximally enhanced. This optimal range corresponds to an interface morphology that consists of well-separated particles which can be treated as individual scattering elements. Hence, the PE increases approximately linearly with increasing $\phi$ until the optimal range is reached. In this regime the increases in PE and the scattering efficiency with increasing particle volume follow the same trend. Further increase in particle concentration beyond the optimal range results in particle aggregation into clusters and strings, both of which red shift and narrows the PE spectrum, causing a decrease in the overall PE.

Our approach leverages the extensive knowledge base that exists on the chemical synthesis of NPs with admirable control on their size, shape and architecture [33]. Once NPs with tailored optical properties are synthesized, spin-coating offers a cost-effective and robust means to transfer them from solution to the substrate. Evaporation of the solvent occurs at room temperature, thus obviating additional energy-intensive steps. This simple nanomanufacturing route to creating plasmonic interfaces on SOI devices presented in this paper paves way to future research to explore the effect of evanescent wave source propagation as the particle distance from the surface could be controlled through shell thickness. Finally, Ag NPs coatings can provide relatively inexpensive means to broadband light harvesting especially if the resulting PE is sufficiently large to offset the material ((~$1/m$^2$) and manufacturing costs. This study clearly demonstrates that by tailoring the interface morphologies through optimizing particle and process parameters the PE could be tripled, a finding that motivates future translational research on integration of such plasmonic layers into thin film Si architectures.

**Acknowledgements:**

We gratefully acknowledge financial support from the National Science Foundation through grants CBET 1049454 and ICORPS 124248. We are thankful to Professor Eric Schiff and Mr. Hui Zhao for providing access to the photocurrent measurement facility, Professor Mathew Maye for the use of the DLS equipment and Professor Amit Agrawal for valuable discussions about this work and help in preparing Figure 1(a). The SOI device fabrication was performed at the Cornell NanoScale Facility, a member of the National Nanotechnology Infrastructure Network, which is supported by the National Science Foundation (grant ECCS 0335765).

Syracuse University has filed a provisional patent application based on the findings of this work.

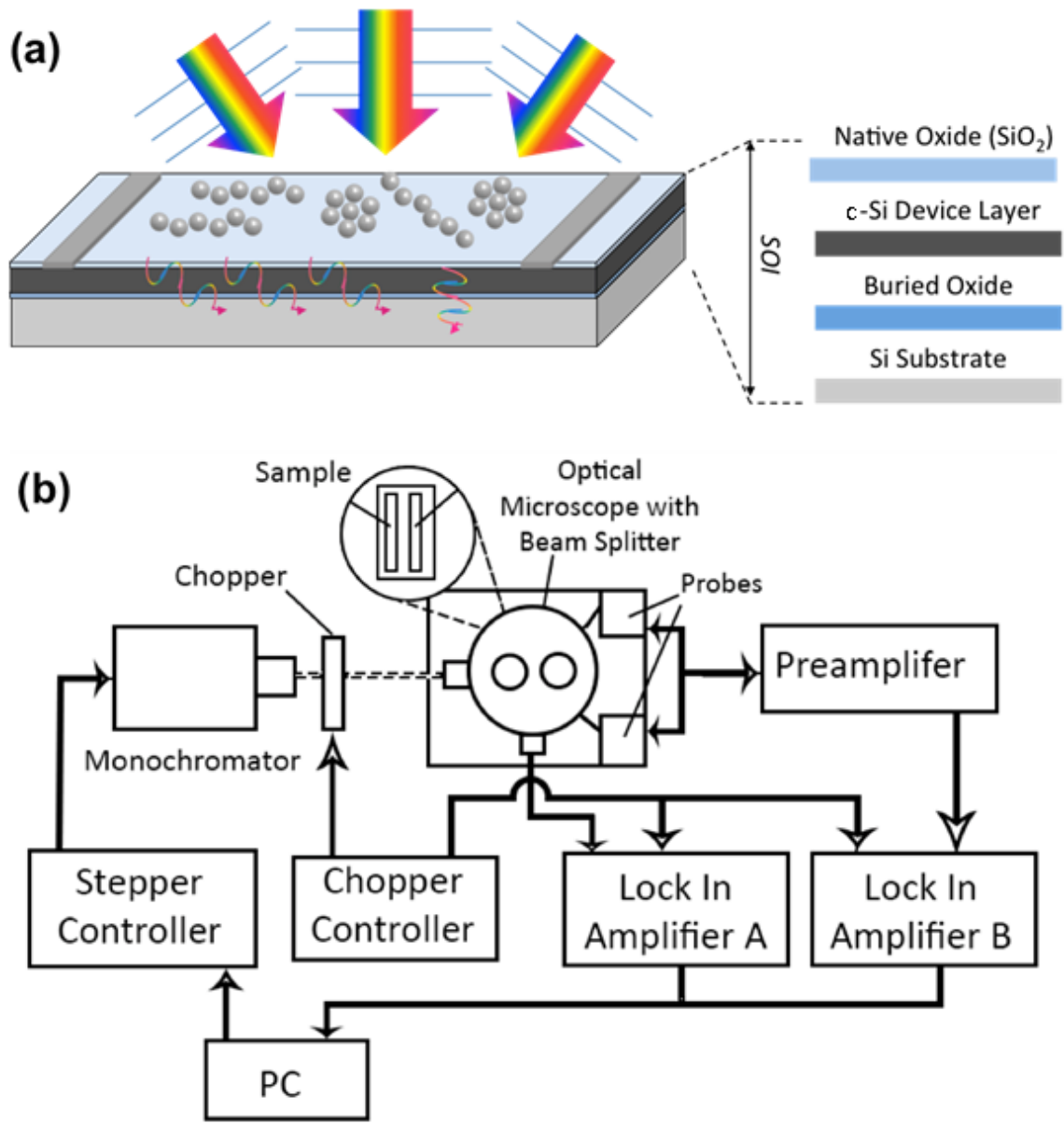

Figure 1: (a) Schematic of the SOI device with nanoparticles. The aluminum contacts were deposited using thermal evaporation. The nanoparticles were deposited through spin-coating. (b) Experimental current collector schematic.

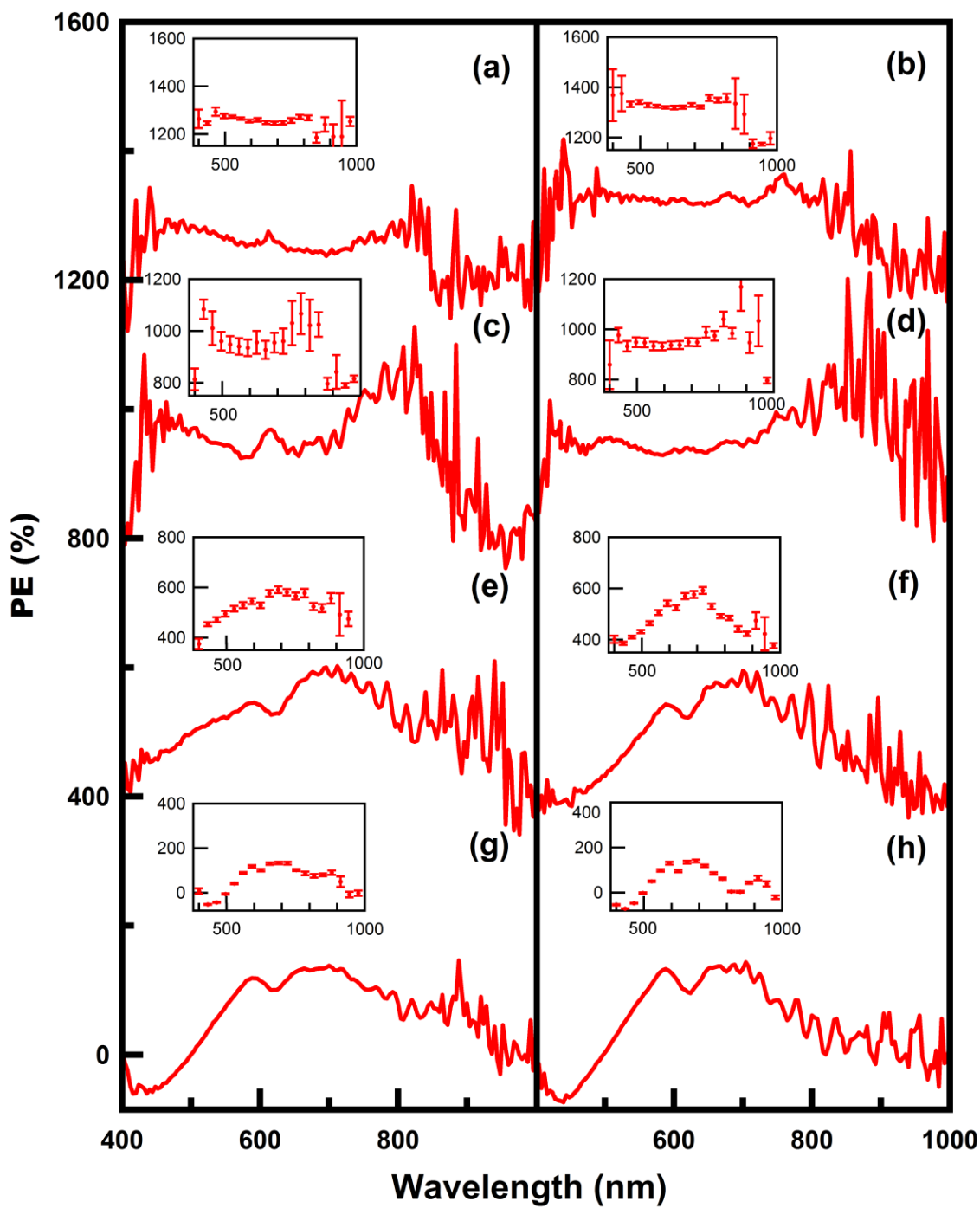

Figure 2: PE response of (a) 0.01%, (b) 0.05%, (c) 0.1%, (d) 0.2%, (e) 0.5%, (f) 1%, (g) 5%, and (h) 10% w/v. Ag nano ink coatings. The graphic inset error bars show a sampling of the standard error for the measurements. Photocurrent measurements are offset by 400 for clarity of display.

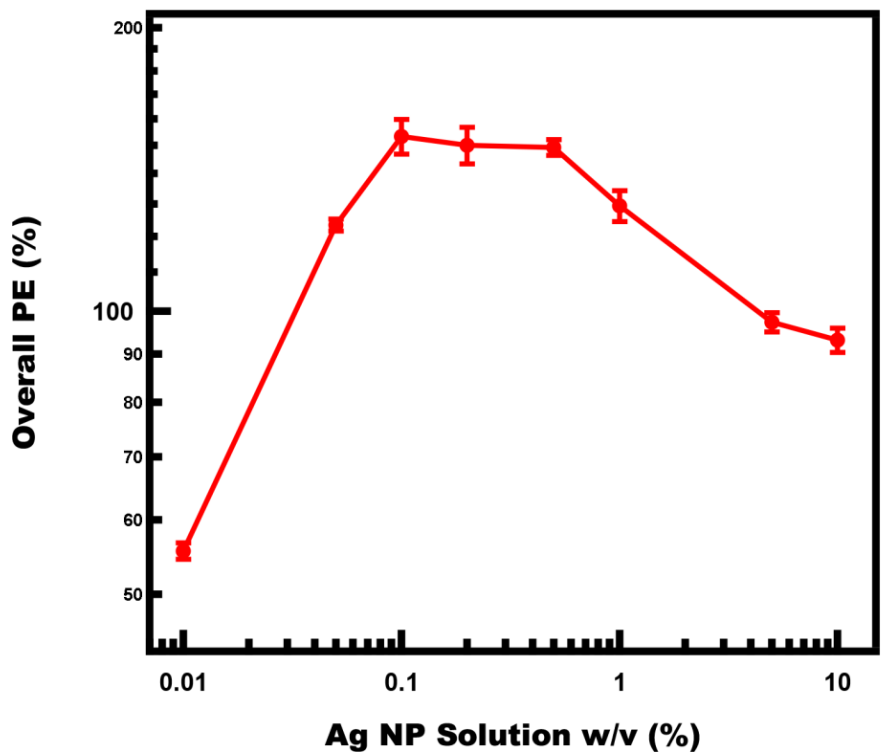

Figure 3: Silver ink solution w/v vs. overall PE. Error bars show the standard error for the measurements.

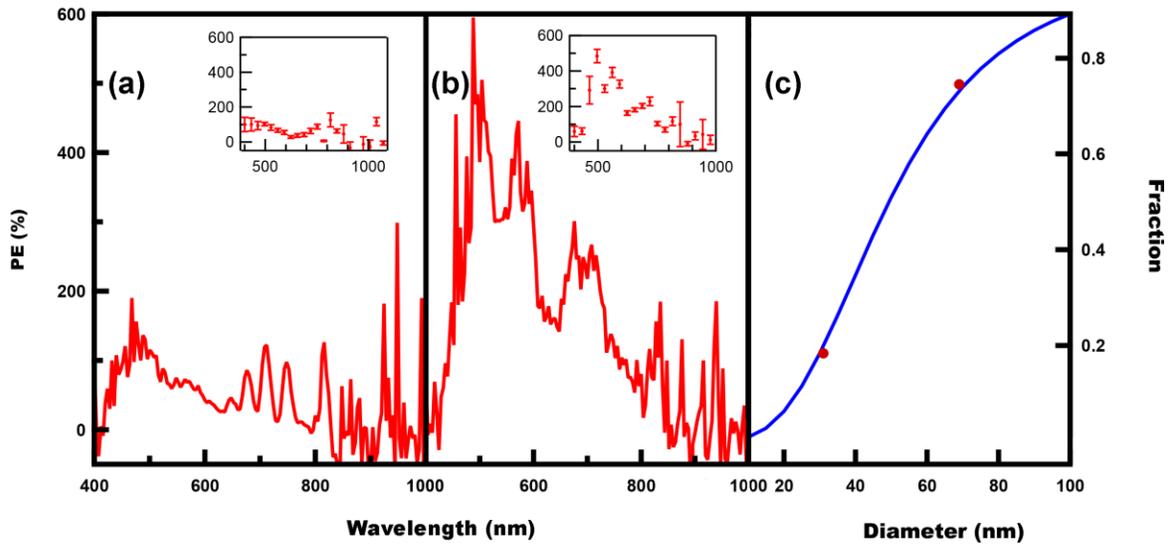

Figure 4: PE response of (a) 31 nm and (b) 69 nm synthesized silver particle coatings. Graphic insets of error bars show the standard error for the measurements. (c) Albedo of particles at 500 nm (blue line) and normalized PE responses for 31, 40, and 69 nm diameters (red circle).

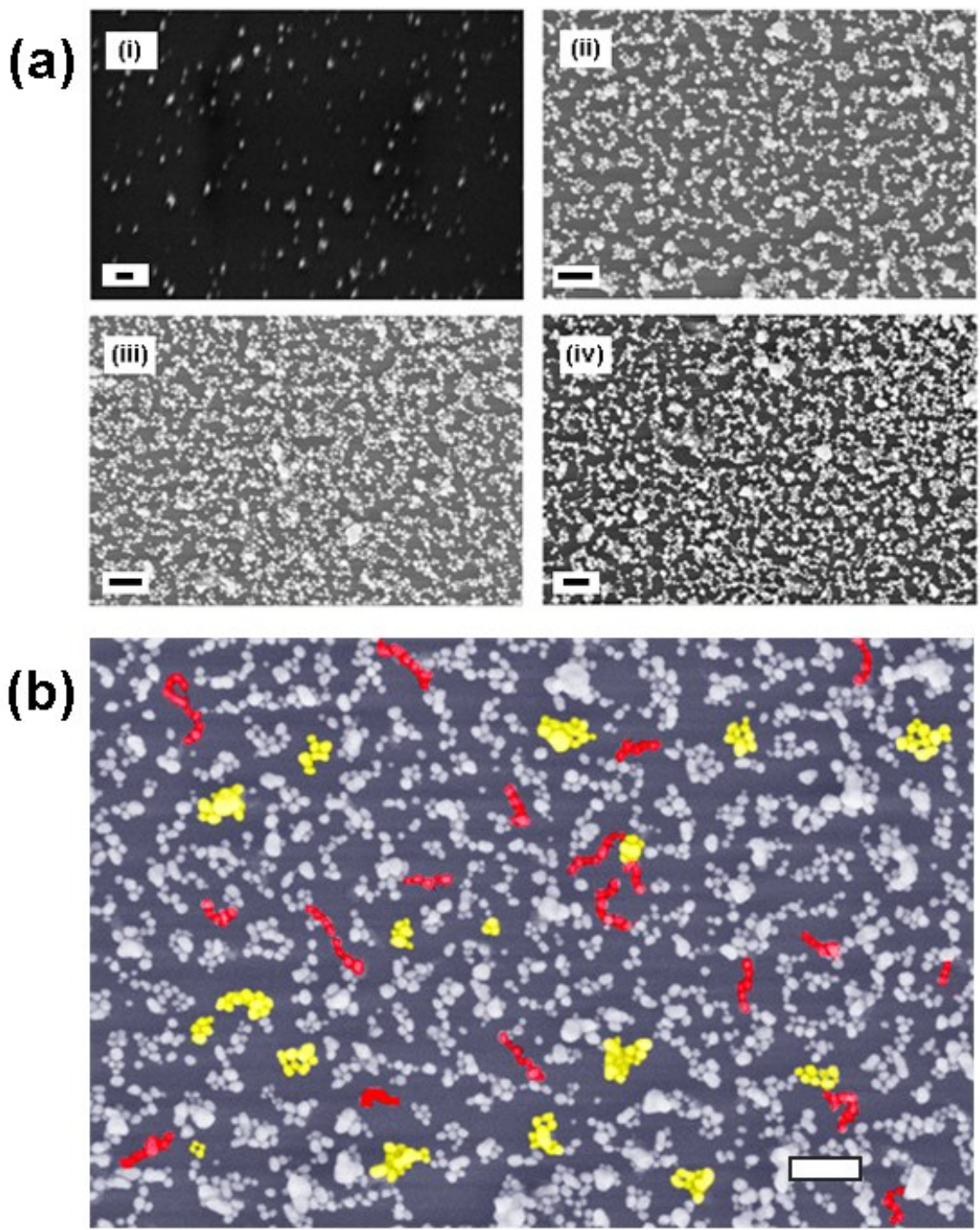

Figure 5: (a) SEM pictures of (i) 0.1%, (ii) 1%, (iii) 5%, and (iv) 10% w/v solutions on SOI wafer. (b) SEM of 1% w/v Ag solution on SOI with example strings (red) and clusters (yellow) colorized. All scale bars are 200 nm.

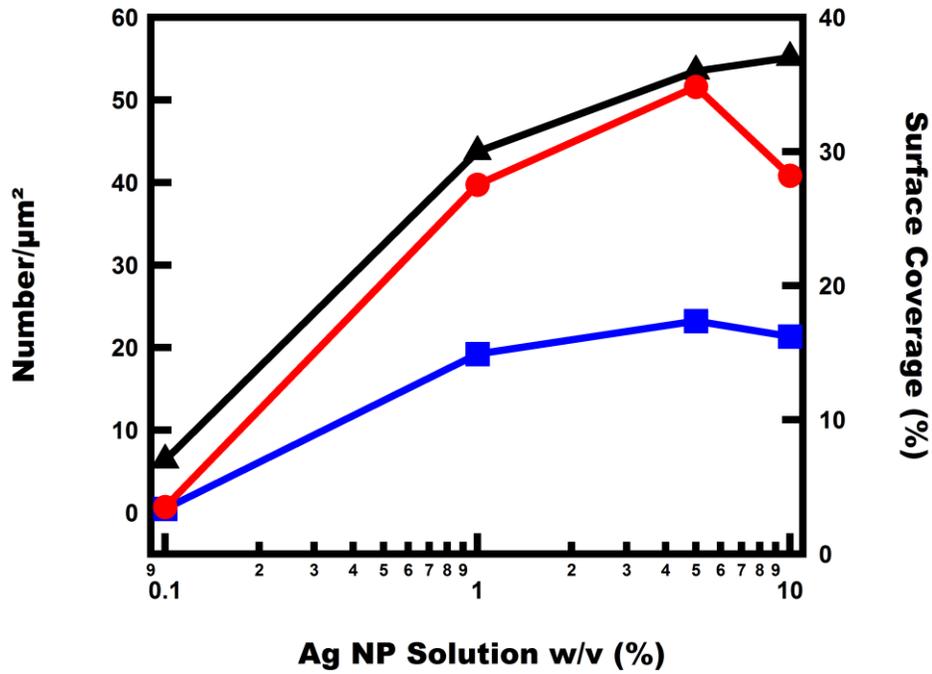

**Figure 6:** Silver solution w/v vs. the 40 nm particle surface coverage (black line) and number of particle formations. The number of clusters (red line) and the number of strings (blue line) are shown per μm$^2$.

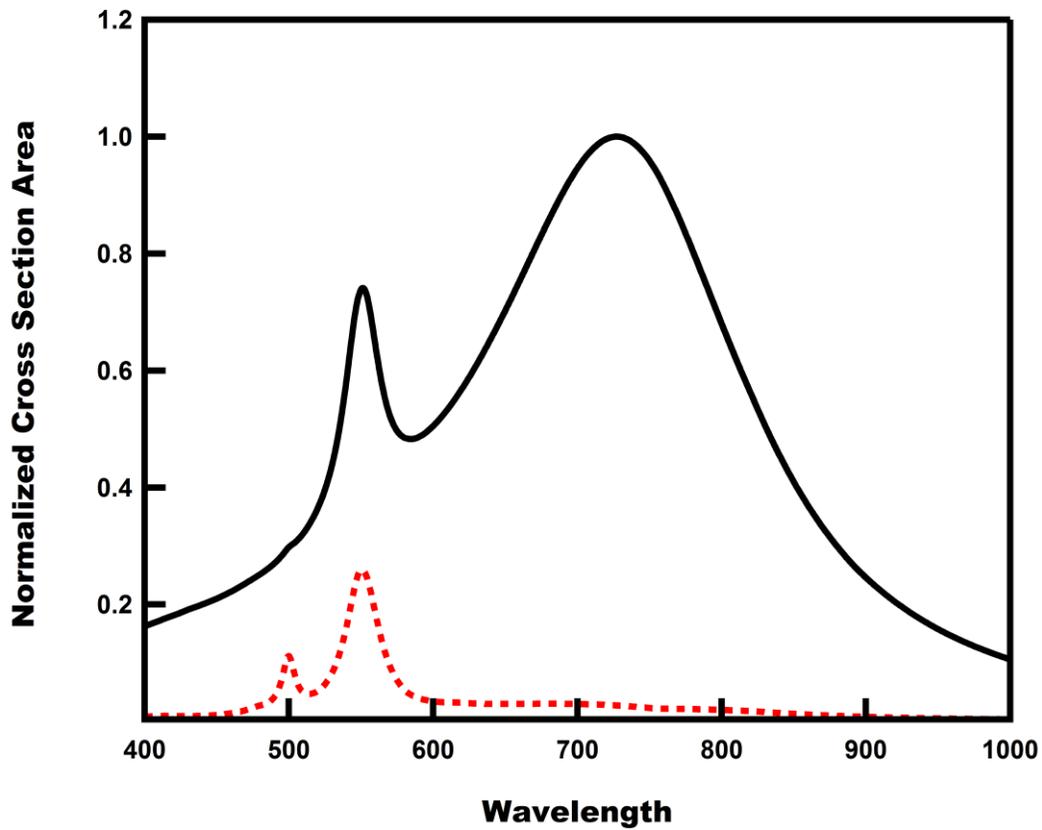

**Figure 7: Mie theory predictions for the scattering (black) and absorption (red) cross sectional areas normalized with respect to the maximum scattering cross sectional area for a 69 nm Ag particle embedded in glucose.**